\documentclass[sigconf, nonacm]{acmart}
\AtBeginDocument{%
  }

\usepackage{multirow}
\usepackage{adjustbox}
\usepackage{bm}

\begin{document}
\sloppy

\title{HSTU-BLaIR: Lightweight Contrastive Text Embedding for Generative Recommender}

% "authornote" and "authornotemark" commands
% used to denote shared contribution to the research.
\author{Yijun Liu}
\email{yijunl@usc.edu}
\affiliation{%
  \institution{Ming Hsieh Department of Electrical and Computer Engineering\\University of Southern California}
  \city{Los Angeles}
  \state{CA}
  \country{United States}
}

\begin{abstract}
Recent advances in recommender systems have underscored the complementary strengths of generative modeling and pretrained language models. We propose HSTU-BLaIR, a hybrid framework that augments the Hierarchical Sequential Transduction Unit (HSTU)-based generative recommender with BLaIR, a lightweight contrastive text embedding model. This integration enriches item representations with semantic signals from textual metadata while preserving HSTU’s powerful sequence modeling capabilities.

We evaluate HSTU-BLaIR on two e-commerce datasets: three subsets from the Amazon Reviews 2023 dataset and the Steam dataset. We compare its performance against both the original HSTU-based recommender and a variant augmented with embeddings from OpenAI’s state-of-the-art \texttt{text-embedding-3-large} model. Despite the latter being trained on a substantially larger corpus with significantly more parameters, our lightweight BLaIR-enhanced approach—pretrained on domain-specific data—achieves better performance in nearly all cases. Specifically, HSTU-BLaIR outperforms the OpenAI embedding-based variant on all but one metric, where it is marginally lower, and matches it on another. These findings highlight the effectiveness of contrastive text embeddings in compute-efficient recommendation settings. Code is available at \url{https://github.com/snapfinger/HSTU-BLaIR}.
\end{abstract}



\maketitle

\section{Introduction}
Recent progress in recommender systems has been fueled by advances in deep learning~\cite{covington2016deep}, and increasingly by techniques from natural language processing~\cite{p5, lyu2023llm, bao2023tallrec, hou2024bridging, hou2024large, notellm, openp5, liao2024llara}. While the dominant industrial paradigm has long relied on multi-stage recommendation pipelines—where two-tower architectures perform efficient candidate retrieval from a massive corpus, followed by a ranking model, often an MLP, for fine-grained scoring~\cite{covington2016deep, yi2019sampling, huang2020embedding, yao2021self}—these systems often require extensive feature engineering and may struggle to capture complex temporal and semantic dependencies. In contrast, generative recommenders jointly model user interaction sequences and directly generate item predictions in an autoregressive fashion~\cite{yuan2019simple, rajput2023gr, zhai2024hstu}. This paradigm enables richer temporal modeling and improved personalization. A key subtask within this space is sequential recommendation, which focuses on capturing the temporal dynamics of user interactions \cite{sasrec, bert4rec}. 

Although recent work has begun to explore the integration of generative recommenders with pretrained language models, most existing approaches rely on prompting or fine-tuning large language models (LLMs)~\cite{p5, gpt4rec, genrec}. These models are often not specifically tailored for recommendation tasks and tend to be computationally inefficient—especially given that state-of-the-art LLMs typically contain billions of parameters~\cite{xu2025slmrec}. As a result, the broader incorporation of pretrained language models into generative recommendation systems remains in its early stages, with critical challenges around scalability, domain adaptation, and efficient utilization of textual signals still largely unresolved.

On the other hand, self-supervised contrastive learning has shown strong potential for producing high-quality text embeddings that capture fine-grained semantic alignment~\cite{gao2021simcse, hou2024bridging}. Generative recommenders, in parallel, excel at modeling structured user behavior over time. Combining these complementary strengths offers a promising direction to unify semantic understanding with behavioral dynamics, enabling more expressive, adaptable, and semantically informed recommendation systems.

In this work, we integrate two recent state-of-the-art models: BLaIR~\cite{hou2024bridging}, a contrastive text encoder pretrained on user reviews and item metadata from the Amazon Reviews 2023 dataset, and HSTU~\cite{zhai2024hstu}, a generative model that formulates recommendation as sequential transduction, unifying retrieval and ranking through autoregressive modeling. BLaIR generates semantically rich textual representations from item metadata, while HSTU provides a scalable, end-to-end architecture for sequential recommendation. By fusing BLaIR’s semantic embeddings with HSTU’s behavioral modeling, our framework enhances personalization while maintaining scalability and interpretability—leveraging human-readable metadata and avoiding the computational overhead of large language models.

\section{Method}

\subsection{Base Model: Hierarchical Sequential Transduction Unit (HSTU)}
We build on the generative recommender of \citet{zhai2024hstu}, which encodes a user’s interaction history with a \textit{Hierarchical Sequential Transduction Unit} (HSTU).  
Stacked Transformer layers with hierarchical attention and relative positional bias map the sequence into autoregressive hidden states, which are subsequently scored against candidate items by a joint retrieval–ranking interaction module.  
We apply its lightweight base configuration, four transformer blocks with four attention heads, to balance efficiency and capacity.
\subsection{Semantic Item Enrichment with \texorpdfstring{BLaIR\textsubscript{BASE}}{BLaIR\_BASE}}
Each item is paired with a fixed textual embedding taken from the publicly released \mbox{BLaIR\textsubscript{BASE}} checkpoint\footnote{\url{https://huggingface.co/hyp1231/blair-roberta-base}} (125 M parameters).  
BLaIR is contrastively pretrained on the first~80 \% of Amazon reviews (timestamp-sorted), ensuring strict temporal separation from our evaluation sets~\cite{hou2024bridging}.

\begin{figure}[t]
    \centering
    \includegraphics[width=\columnwidth]{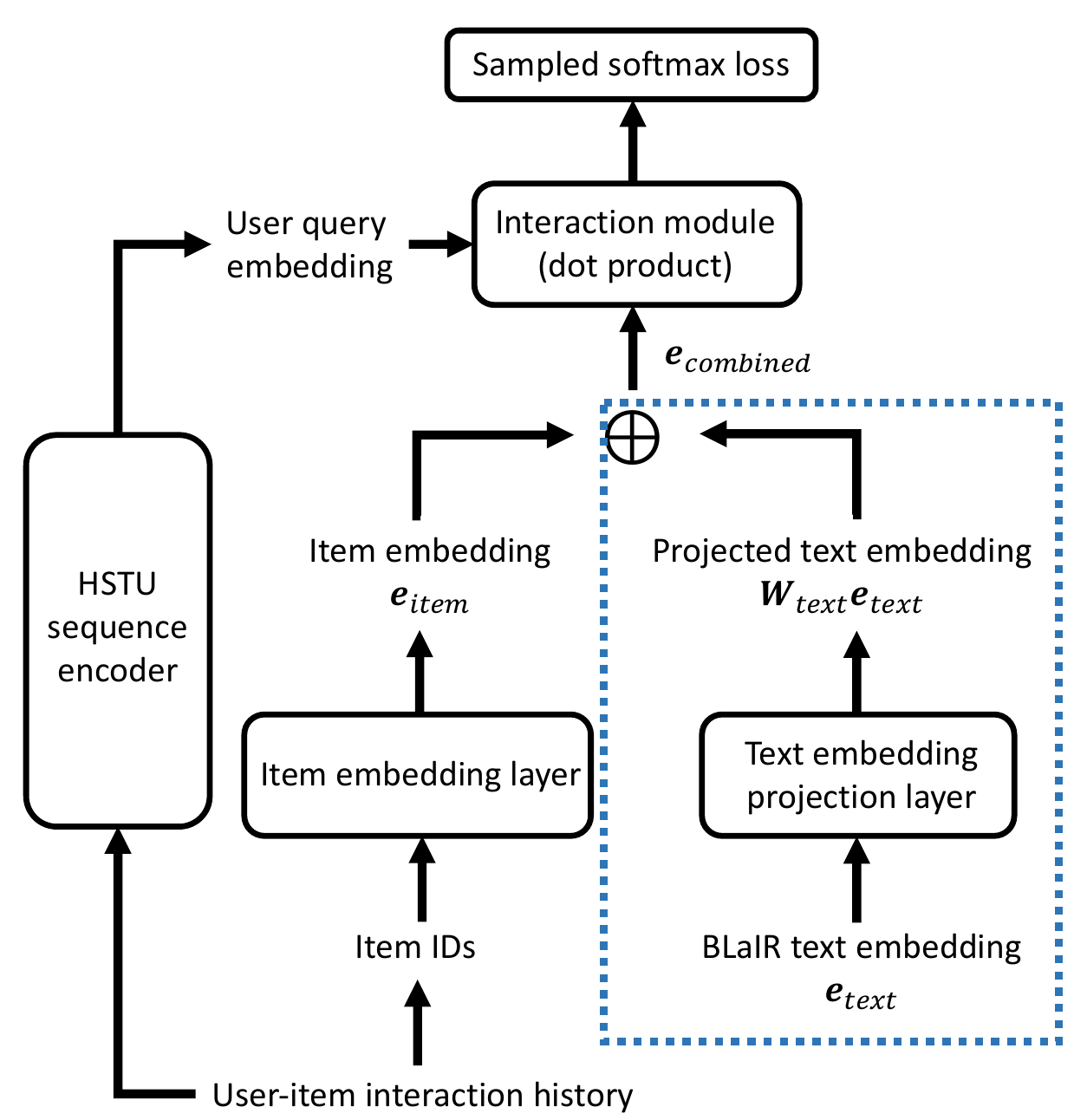}
    \caption{HSTU-BLaIR architecture.  Pre-computed BLaIR embeddings are projected and fused with trainable ID embeddings (\(\oplus\)).  The dotted box highlights the fusion module.}
    \label{fig:pipeline}
\end{figure}

Let \(\mathbf{e}_{\text{id}}\!\in\!\mathbb{R}^{d}\) be the trainable ID embedding of an item and  
\(\mathbf{e}_{\text{text}}\!\in\!\mathbb{R}^{d_t}\) its corresponding BLaIR textual embedding.  
A learnable projection \(\mathbf{W}_{\text{text}}\!\in\!\mathbb{R}^{d\times d_t}\) maps textual features into the ID-embedding space, after which element-wise addition yields a unified representation
\begin{equation}
    \mathbf{e}_{\text{combined}}
    \;=\;
    \mathbf{e}_{\text{id}}
    \;+\;
    \textbf{W}_{\text{text}}\mathbf{e}_{\text{text}}
\end{equation}
The fused vector is combined with positional encodings and dropout before entering the HSTU encoder, preserving dimensionality while injecting semantic context (Figure~\ref{fig:pipeline}).

\subsection{Text-Aware Local Negative Sampling}
To supply semantically informed negatives, we extend the original local negative sampler by augmenting each sampled item with text embedding while leaving the sampling distribution unchanged.  
Specifically, for any sampled item the returned embedding is
\begin{equation}
    \mathbf{e}_{\text{sampled}}
    \;=\;
    \frac{
        \mathbf{e}_{\text{id}} + \mathbf{W}_{\text{neg}}\mathbf{e}_{\text{text}}
    }{
        \bigl\lVert \mathbf{e}_{\text{id}} + \mathbf{W}_{\text{neg}}\mathbf{e}_{\text{text}} \bigr\rVert_{2}
    }
\end{equation}
where \(\mathbf{W}_{\text{neg}}\!\in\!\mathbb{R}^{d\times d_t}\) is a projection matrix dedicated to the sampler.  
This modification injects semantic structure into the contrastive loss without altering the uniform ID-selection strategy or incurring additional sampling overhead.

\subsection{Datasets}

\paragraph{\textbf{Amazon Reviews 2023 Dataset}}  
We train and evaluate HSTU-BLaIR on the \textit{Video Games}, \textit{Office Products}, and \textit{Musical Instruments} subsets of the Amazon Reviews 2023 dataset~\cite{hou2024bridging}. Each subset contains timestamped user purchase histories, product metadata (e.g., title and category), and review texts. We use the 5-core version of the dataset, filtering out users and items with fewer than five interactions, following the offline evaluation protocol in~\cite{zhai2024hstu}.

\paragraph{\textbf{Steam Dataset}}  
To evaluate the generalization ability of our approach beyond traditional retail platforms, we further experiment on the Steam dataset~\cite{sasrec}, which represents a digital goods marketplace. We construct each item’s metadata by concatenating its app name, publisher, developer, genres, tags, and specifications into a single input sentence. User histories are constructed based on the user's review timestamp, and we filter out users with fewer than five interactions and discarding items without metadata.

Steam is entirely disjoint from the Amazon corpus and was not seen by the BLaIR encoder during pretraining, thus serving as a robust cross-domain benchmark that eliminates potential overlap. Dataset statistics are summarized in Table~\ref{tab:dataset_stats}.

\begin{table}[h]
  \centering
  \resizebox{\linewidth}{!}{%
    \begin{tabular}{lccc}
      \hline
      & \textbf{Items} & \textbf{Users} & \textbf{Interactions} \\
      \hline
      \textbf{Video Games}      & 25,612   & 94,762   & 814,585   \\
      \textbf{Office Products}  & 77,551   & 223,308  & 1,800,877 \\
      \textbf{Musical Instruments}         & 24,587   & 57,439   & 511,835   \\
      \textbf{Steam}            & 11,808   & 139,259   & 1,538,285   \\
      \hline
    \end{tabular}
  }
    \caption{Statistics of the three Amazon Reviews 2023 subsets and the Steam dataset used in our sequential recommendation experiments, including the number of users, items, and interactions.}
  \label{tab:dataset_stats}
\end{table}

\begin{table*}[t]
\centering
\caption{Evaluation metrics on the Video Games, Office Products, and Musical Instruments subsets of the Amazon Reviews 2023 dataset, as well as the Steam dataset. Each cell shows absolute values (top row) and percentage improvements over HSTU / SASRec (bottom row). Best values per column are bolded.}
\adjustbox{max width=\textwidth}{
\begin{tabular}{clcccccc}
\toprule
Dataset & Model & HR@10 & HR@50 & HR@200 & NDCG@10 & NDCG@200 & MRR \\
\midrule
\multirow{8}{*}{\shortstack[c]{\large Video\\\large Games}}
& SASRec              & .1028 & .2317 & .3941 & .0573 & .1097 & .0518 \\
&                     & — & — & — & — & — & — \\
& HSTU                & .1315 & .2765 & .4565 & .0741 & .1327 & .0658 \\
&                     & (+28\%) & (+19.3\%) & (+15.8\%) & (+29.3\%) & (+21.0\%) & (+27.1\%) \\
& HSTU-OpenAI (TE3L)  & .1328 & .2821 & .4645 & .0742 & .1341 & .0658 \\
&                     & (+1.0\% / +29.2\%) & (+2.0\% / +21.8\%) & (+1.8\% / +17.9\%) & (+0.1\% / +29.5\%) & (+1.1\% / +22.2\%) & (0.0\% / +27.0\%) \\
& \textbf{HSTU-BLaIR} & \textbf{.1353} & \textbf{.2852} & \textbf{.4684} & \textbf{.0760} & \textbf{.1361} & \textbf{.0674} \\
&                     & (+2.9\% / +31.6\%) & (+3.1\% / +23.1\%) & (+2.6\% / +18.9\%) & (+2.6\% / +32.6\%) & (+2.6\% / +24.1\%) & (+2.4\% / +30.1\%) \\
\midrule
\multirow{8}{*}{\shortstack[c]{\large Office\\\large Products}}
& SASRec              & .0281 & .0668 & .1331 & .0153 & .0335 & .0143 \\
&                     & — & — & — & — & — & — \\
& HSTU                & .0395 & .0880 & .1649 & .0223 & .0443 & .0207 \\
&                     & (+40.6\%) & (+31.7\%) & (+23.9\%) & (+45.8\%) & (+32.2\%) & (+44.8\%) \\
& HSTU-OpenAI (TE3L)  & .0477 & .1050 & .1940 & .0269 & .0526 & .0247 \\
&                     & (+20.8\% / +69.8\%) & (+19.3\% / +57.2\%) & (+17.6\% / +45.8\%) & (+20.6\% / +75.8\%) & (+18.7\% / +57.0\%) & (+19.3\% / +72.7\%) \\
& \textbf{HSTU-BLaIR} & \textbf{.0484} & \textbf{.1068} & \textbf{.1946} & \textbf{.0271} & \textbf{.0529} & \textbf{.0248} \\
&                     & (+22.5\% / +72.2\%) & (+21.4\% / +59.9\%) & (+18.0\% / +46.2\%) & (+21.5\% / +77.1\%) & (+19.4\% / +57.9\%) & (+19.8\% / +73.4\%) \\
\midrule
\multirow{8}{*}{\shortstack[c]{\large Musical\\\large Instruments}}
& SASRec              & .0643 & .1488 & .2784 & .0356 & .0731 & .0326 \\
&                     & — & — & — & — & — & — \\
& HSTU                & .0700 & .1599 & .2910 & .0392 & .0783 & .0359 \\
&                     & (+8.9\%) & (+7.5\%) & (+4.5\%) & (+10.1\%) & (+7.1\%) & (+10.1\%) \\
& HSTU-OpenAI (TE3L)  & .0708 & .1635 & .3005 & .0393 & .0798 & .0360 \\
&                     & (+1.1\% / +10.1\%) & (+2.3\% / +9.9\%) & (+3.3\% / +7.9\%) & (+0.3\% / +10.4\%) & (+1.9\% / +9.2\%) & (+0.3\% / +10.4\%) \\
& \textbf{HSTU-BLaIR} & \textbf{.0733} & \textbf{.1681} & \textbf{.3066} & \textbf{.0406} & \textbf{.0818} & \textbf{.0371} \\
&                     & (+4.7\% / +14.0\%) & (+5.1\% / +13.0\%) & (+5.4\% / +10.1\%) & (+3.6\% / +14.0\%) & (+4.5\% / +11.9\%) & (+3.3\% / +13.8\%) \\
\midrule
\midrule
\multirow{8}{*}{\shortstack[c]{\large Steam}}
& SASRec              & .0881 & .2277 & .4312 & .0462 & .1068 & .0426 \\
&                     & — & — & — & — & — & — \\
& HSTU                & .1038 & .2575 & .4704 & .0544 & .1195 & .0492 \\
&                     & (+17.8\%) & (+13.1\%) & (+9.1\%) & (+17.8\%) & (+11.9\%) & (+15.5\%) \\
& HSTU-OpenAI (TE3L)  & .1089 & .2657 & .4806 & .0579 & \textbf{.1241} & \textbf{.0525} \\
&                     & (+4.9\% / +23.6\%) & (+3.2\% / +16.7\%) & (+2.2\% / +11.5\%) & (+6.4\% / +25.3\%) & (+3.8\% / +16.2\%) & (+6.7\% / +23.2\%) \\
& \textbf{HSTU-BLaIR} & \textbf{.1100} & \textbf{.2668} & \textbf{.4812} & \textbf{.0581} & \textbf{.1241} & .0523 \\
&                     & (+6.0\% / +24.9\%) & (+3.6\% / +17.2\%) & (+2.3\% / +11.6\%) & (+6.8\% / +25.8\%) & (+3.8\% / +16.2\%) & (+6.3\% / +22.8\%) \\
\bottomrule
\end{tabular}
}
\label{tab:result}
\end{table*}

\subsection{Tasks}
We evaluate our model on next-item prediction in a sequential recommendation setting. Following \citet{zhai2024hstu}, user interactions are sorted chronologically, and a leave-one-out evaluation strategy is applied: for each user, the most recent interaction is held out for testing, while all prior interactions are used for training. We also adopt the training and evaluation protocol from \citet{zhai2024hstu}, which includes full data shuffling and multi-epoch training. All models, including our HSTU-BLaIR and the baselines, are trained for 100 epochs on the Amazon Reviews 2023 dataset as well as the Steam dataset.

\subsection{Baselines}

We compare our proposed framework against the following baselines:

\begin{enumerate}
    \item \textbf{SASRec}~\cite{sasrec}, a widely used transformer-based sequential recommender that models item sequences using self-attention over ID embeddings;
    
    \item \textbf{HSTU}~\cite{zhai2024hstu}, the original generative recommender, which encodes user histories using a hierarchical transformer architecture with only learned item ID embeddings and no textual input;
    
    \item \textbf{HSTU-OpenAI (TE3L)}, a variant of HSTU-based GR in which item textual embeddings are derived from OpenAI’s \texttt{text-embedding-3-large} model~\cite{openai2024embedding}, serving as an alternative to the BLaIR embeddings.
\end{enumerate}

While OpenAI has not released architectural or training details for \texttt{text-embedding-3-large}, prior disclosures suggest it likely contains billions of parameters and was trained on broad corpora including Wikipedia, books, and large-scale web data~\cite{brown2020language}. In contrast, BLaIR\textsubscript{BASE}~\cite{hou2024bridging} is a domain-specific contrastive model with only 125M parameters, pretrained on Amazon Reviews 2023.

Notably, \texttt{text-embedding-3-large} returns unit-normed embeddings by default, whereas BLaIR imposes no explicit normalization constraint. We preserve the native output behavior of each encoder to reflect realistic usage and maintain their respective representational properties.

The \texttt{text-embedding-3-large} model supports a maximum input length of 8,179 tokens. In our datasets, only a small number of items—e.g., 2 in Video Games and 4 in Office Products—exceed this limit. For such cases, we retain only the first 8,179 tokens. In contrast, BLaIR was pretrained with a much shorter context window of 64 tokens. However, following the official usage example provided on its Hugging Face model page,\footnote{\url{https://huggingface.co/hyp1231/blair-roberta-base}} we tokenize item texts using a maximum length of 512 tokens. While this exceeds the model's original pretraining window, we found it performs robustly in practice.

\subsection{Evaluation Metrics}

We assess model performance using two standard ranking metrics commonly used in recommendation systems:

\textbf{HR@K (Hit Rate at K)} measures whether the ground-truth next item appears within the top-$K$ predicted items. It reflects the model’s ability to include the correct item in its top-$K$ recommendations, regardless of position.

\textbf{NDCG@K (Normalized Discounted Cumulative Gain at K)} extends this by considering the position of the correct item in the ranked list, assigning higher scores to items that appear earlier. It is defined as $\text{NDCG@}K = \frac{\text{DCG@}K}{\text{IDCG@}K}$, where $\text{DCG@}K = \sum_{i=1}^{K} \frac{rel_i}{\log_2(i+1)}$ and $\text{IDCG@}K$ denotes the maximum possible DCG for an ideal ranking. Here, $rel_i$ is a binary relevance indicator (1 if the item at position $i$ is correct, 0 otherwise).

\textbf{MRR (Mean Reciprocal Rank)} computes the average of the reciprocal ranks of the first relevant item across all queries. It is defined as $\text{MRR} = \frac{1}{N} \sum_{i=1}^{N} \frac{1}{\text{rank}_i}$, where $\text{rank}_i$ is the position of the first relevant item for the $i$-th query. MRR rewards models that rank the correct item higher and is useful for single-ground-truth settings.

\section{Results}

Table~\ref{tab:result} reports the performance of our proposed HSTU-BLaIR method in comparison to several baselines across four benchmarks: the Video Games, Office Products, and Musical Instruments subsets of the Amazon Reviews 2023 dataset, as well as the Steam dataset. As shown in Table~\ref{tab:dataset_stats}, these datasets vary in scale and sparsity, with Office Products featuring a significantly larger user and item set, resulting in a sparser interaction matrix.

Across all datasets and evaluation metrics, HSTU-BLaIR consistently outperforms the original HSTU model, demonstrating the effectiveness of incorporating semantic item information via BLaIR’s pretrained textual embeddings. Compared to HSTU-OpenAI (TE3L), which leverages OpenAI’s \texttt{text-embedding-3-large}, HSTU-BLaIR achieves stronger performance on the majority of metrics. Specifically, it surpasses the OpenAI variant on 22 of 24 comparisons across datasets, with one metric slightly lower (MRR on Steam) and one metric equal (NDCG@200 on Steam). These results suggest that domain-specific contrastive encoders like BLaIR can provide more effective and compute-efficient representations for recommendation than general-purpose large language models.

Improvements are especially pronounced in the sparser Office Products dataset, where HSTU-BLaIR achieves up to a 77\% gain in NDCG@10 over SASRec, and up to 21.5\% over HSTU. These results highlight the effectiveness and generalizability of our approach across different dataset scales and sparsity levels.

\section{Conclusion and Future Work}

This work investigates the integration of contrastively learned textual embeddings into a generative sequential recommender system. Our findings demonstrate that enriching user sequence models with semantically grounded text representations yields consistent improvements in recommendation performance across multiple benchmarks. Specifically, we combine BLaIR-generated text embeddings with HSTU’s item ID embeddings via element-wise addition, allowing the model to jointly leverage both structured (ID-based) and unstructured (textual) item information.

To explore more flexible fusion strategies, we also implemented a learnable vector gating mechanism, where each embedding dimension independently balances ID and textual contributions. Surprisingly, this approach consistently underperformed compared to simple addition, likely due to optimization instability or insufficient signal early in training. As future work, we plan to investigate context-aware fusion strategies, such as user-conditioned or attention-based gating, as well as fine-tuning the text encoder to improve alignment with downstream objectives.

\section*{Limitations}
While our approach demonstrates the effectiveness of integrating contrastive textual signals into generative sequential recommenders, it has several limitations.

First, the BLaIR text embeddings are kept frozen during training. While this design improves efficiency and stability, it restricts the model’s ability to adapt text features to the specific downstream task. Second, the fusion of item ID embeddings and textual representations is relatively shallow (e.g., element-wise summation), which may limit the capacity of the model to capture rich interactions between the two modalities. Finally, the effectiveness of BLaIR embeddings depends heavily on the availability and quality of item metadata. In domains with sparse, noisy, or inconsistent textual content, the benefits of leveraging pretrained text representations may diminish.

\bibliographystyle{ACM-Reference-Format}
\bibliography{KDD25_bib}

% %%
% %% If your work has an appendix, this is the place to put it.
% \appendix



\end{document}